# Excitonic effect on two-photon absorption of two-dimensional semiconductors: Theory and applications to MoS$_2$ and WS$_2$ monolayers

You-Zhao Lan*[1] and Xiao-Hu Bao

*Institute of Physical Chemistry, Key Laboratory of the Ministry of Education for Advanced Catalysis Materials, College of Chemistry and Life Sciences, Zhejiang Normal University, Zhejiang, Jinhua 321004, China*

**Abstract**

Based on the Bethe-Salpeter equation eigenstates, we present a first-principles many-body formalism for calculating the two-photon absorption (TPA) coefficient of semiconductors. We apply this formalism to calculate the TPA spectra of MoS$_2$ and WS$_2$ monolayers. The all-electron full-potential linearised augmented-plane wave based functions are used for solving the Bethe-Salpeter equation. The calculated spectra are in good agreement with the available experimental ones for WS$_2$ monolayer. The calculated TPA spectra exhibit significant excitonic effects when compared to those based on the independent particle approximation. The physical origin of TPA excitonic transitions of MoS$_2$ and WS$_2$ monolayers are revealed by tracing the sum-over-states process. We show that the spin-orbit coupling effect leads to characteristic double peaks with an interval of half spin-orbit splitting energy. These double peaks mainly originate from the transitions at the vicinity of K point.

1. Introduction

It has been well known that the excitonic effect must be considered to accurately understand the one-photon absorption (OPA) spectrum of materials [1–9]. The excitonic effect includes the electron-hole interaction on the calculated absorption spectrum, which goes beyond the independent particle approximation (IPA). As reported by Rohlfing and Louie [8], with inclusion of the excitonic effect, the calculated OPA spectrum is in excellent agreement with the experimental one in terms of peak positions and strengths. We can expect that considering the excitonic effect should be also necessary to understand the two-photon absorption (TPA) spectrum of materials. The TPA spectroscopy is also an important tool to study the excited state of system. The TPA and OPA spectra generally provide the complementary information for the excited state of system because they have different selection rules. For instance, for a centrosymmetric system, one- and two-photon allowed transitions are mutually exclusive, and thus the TPA spectrum can provide the dark states that do not appear in the OPA spectrum. Similar to OPA, many simple or empirical models based on a few bands within IPA have been developed to understand the TPA of solids [10,11]. Obviously, information provided by these models is limited for the excited state of system because these models miss the possible excitonic state of system.

---

[1] Corresponding author: Youzhao Lan; Postal address: Institute of Physical Chemistry, College of Chemistry and Life Sciences, Zhejiang Normal University, Zhejiang, Jinhua 321004, China; Fax: +086 579 82282269; E-mail address: lyz@zjnu.cn

In this paper, based on the Bethe-Salpeter equation (BSE) eigenstates, we present a first-principles many-body formalism to calculate the TPA spectrum. The strategy is first to solve the self-consistent Kohn–Sham equations with the generalized gradient approximation (GGA) of the Perdew–Burke–Ernzerhof (PBE) functional [12] combined with the all-electron full-potential linearised augmented-plane wave (FP-LAPW) method [13], and then to solve the BSE to obtain the excited states of system, finally based on the BSE eigenstates, we use the time-dependent perturbation theory to obtain an expression for calculating the TPA coefficient. Applications are performed on two monolayer transition-metal dichalcogenides, *i.e.* $MoS_2$ and $WS_2$, whose OPA spectra including the excitonic effect have been well studied [4,5,14,15]. Compared to the IPA-TPA spectra, the BSE-TPA ones exhibit significant excitonic effects. For $WS_2$ monolayer, the BSE-TPA spectrum is in good agreement with the experimental TPA one. At the same time, we also calculate the BSE-OPA spectra of $MoS_2$ and $WS_2$ monolayers, which are in excellent agreement with the very recent experimental results [6] in terms of peak positions and line shape. By tracing the sum-over-states (SOS) process of the TPA coefficients, we discuss the physical origin of the TPA states.

In Section 2, we describe a detailed theoretical strategy for calculating the TPA spectrum. In Section 3, we give the computational details, after which the physical origin of TPA of $MoS_2$ and $WS_2$ monolayers is discussed and a comparison between the theoretical and experimental TPA spectra for $WS_2$ monolayer is made. Finally, the conclusions are given in Section 4.

## 2. Theory

We first sketch out how to obtain the optical transition rate by solving the following time-dependent

Schrödinger equation in the *interaction* picture:

$$i\hbar \frac{\partial \psi(t)}{\partial t} = [H_0 + V(t)]\psi(t) \quad (1)$$

, where $H_0$ and $V(t)$ are the static unperturbed Hamiltonian and the time-dependent perturbation operator (describing the interaction between the light radiation and material), respectively. The $V(t)$ in the interaction picture is defined as

$$V(t) = e^{(i/\hbar)H_0 t} V_0 e^{-i\omega t + (\eta/\hbar)t} e^{-(i/\hbar)H_0 t} \quad (2)$$

, where $V_0$ is a static operator, $\omega$ is an applied field frequency, and $\eta$ is a positive infinitesimal energy. The various order transition rates for a direct optical transition from an initial state $|i\rangle$ to a final state $|f\rangle$ (two eigenstates of $H_0$), accompanied by the simultaneous absorption of $n$ photons (each of frequency, $\omega$), are given by the matrix elements of $V_0$ and the Dirac delta function as [11]

$$W_n(\omega) = \frac{2\pi}{\hbar} \left| \sum_{a_1,a_2,\cdots,a_{n-2},a_{n-1}} \frac{\langle f|V_0|a_{n-1}\rangle\langle a_{n-1}|V_0|a_{n-2}\rangle}{[E_{n-1} - E_{n-2} - (n-1)\hbar\omega]} \cdots \frac{\langle a_2|V_0|a_1\rangle\langle a_1|V_0|i\rangle}{(E_2 - E_1 - 2\hbar\omega)(E_1 - E_i - \hbar\omega)} \right|^2 \delta(E_f - E_i - n\hbar\omega) \quad (3)$$

, where $|a_1\rangle, |a_2\rangle, \ldots, |a_{n-1}\rangle$ are all possible intermediate states, and the corresponding energies are $E_1, E_2, \ldots, E_{n-1}$, respectively. A detailed procedure for deriving the first, second, and third-order optical transition rates (*i.e.*, $n = 1, 2, 3$ in Eq.3) can be found in the textbook [3].

Then, we consider two types of unperturbed Hamiltonians ($H_0$) to calculate the optical transition rate of a bulk system. One is one-particle Hamiltonian ($H_0^{1p}$) within the IPA, the other is effective two-particle Hamiltonian ($H_0^{2p}$) based on the BSE. Within the IPA, the eigenstate and corresponding energy in Eq.3 are taken from the independent particle band structure, that is, $|i\rangle$ and $|f\rangle$ are assumed to be the occupied valence band ($v$) and the unoccupied conduction band ($c$), respectively, and $E_f - E_i$ denotes the transition energy from a valence band ($E_v$) to a conduction band ($E_c$). At this time, the total first and second-order transition rates per unit volume for a bulk system can be written by

$$W_1^{(1)}(\omega) = \frac{4\pi}{\hbar} \frac{1}{\Omega} \sum_{c,v,k} |\langle c|V_0|v\rangle|^2 \delta(E_c - E_v - \hbar\omega) \quad (4)$$

$$W_2^{(2)}(\omega) = \frac{4\pi}{\hbar} \frac{1}{\Omega} \sum_{c,v,k} \left| \sum_{a_1} \frac{\langle c|V_0|a_1\rangle\langle a_1|V_0|v\rangle}{E_1 - E_v - \hbar\omega} \right|^2 \delta(E_c - E_v - 2\hbar\omega) \quad (5)$$

, where a factor of 2 is included for electron-spin degeneracy, $\Omega$ is the volume of unit cell, and the $k$-dependences of eigenstates and their energies are compressed for clarity.

The use of effective two-particle Hamiltonian ($H_0^{2p}$) based on the BSE make the electron-hole interaction be included in the calculation of optical transition rates. The $H_0^{2p}$ is defined as [7,16]

$$H_0^{2p} = H_0^{1p} + H^{eh} \qquad (6)$$

, where $H_0^{1p}$ has the same form as one-particle Hamiltonian based on the IPA, and $H^{eh}$ is the electron-hole interaction term which includes the direct attraction interaction term ($H^{eh,d}$) and the exchange term ($H^{eh,x}$). While the $H_0^{1p}$ describes the independent-particle excitation, the $H^{eh}$ means the coupling between different independent-particle transitions ($v \to c$). In this case, the eigenstates of $H_0^{2p}$ indicate the electron-hole excited states $|S\rangle$ (*i.e.*, $H_0^{2p}|S\rangle = E^S|S\rangle$). The $|S\rangle$ is usually given by the linear combination of independent-particle excitations $|vck\rangle$ (*i.e.*, $|vk\rangle$ to $|ck\rangle$) as

$$|S\rangle = \sum_{c,v,k} A_{c,v,k}^S |vck\rangle \qquad (7)$$

Now, $W_T^{(1)}$ and $W_T^{(2)}$ are written by

$$W_1^{(1)}(\omega) = \frac{4\pi}{\hbar} \sum_{S_f} \left| \langle 0|V_0|S_f\rangle \right|^2 \delta\left(E_f^S - \hbar\omega\right) \qquad (8)$$

$$W_2^{(2)}(\omega) = \frac{4\pi}{\hbar} \sum_{S_f} \left| \sum_m \frac{\langle 0|V_0|S_m\rangle \langle S_m|V_0|S_f\rangle}{E_m^S - \hbar\omega} \right|^2 \delta\left(E_f^S - 2\hbar\omega\right) \qquad (9)$$

, where $|0\rangle$ is the electronic ground state (initial state) and $|S_m\rangle$ and $|S_f\rangle$ are the intermediate and final states of an optical transition with the excitation energies of $E_m^S$ and $E_f^S$, respectively.

For the interaction between light radiation and material, we use $V_0 = (e/mc)\mathbf{A}\cdot\mathbf{p}$ at the momentum gauge, where $\mathbf{A}$ is the vector potential associated with the applied light radiation and $\mathbf{p}$ is the momentum operator for the electron. In this case, Eqs.4 and 5 require the computation of momentum matrix between independent-particle states (*e.g.*, $<c|p|v>$), and Eqs.8 and 9 require $<0|p|S_f>$ and $<S_m|p|S_f>$. As $p$ is a one-particle operator, we can easily compute the $<0|p|S_f>$ and $<S_m|p|S_f>$ in terms of the general rule for the matrix element of one-particle operator between Slater determinants [3,17] and the renormalization of momentum matrix elements of IPA [18–20]. For example,

$$\langle 0|p|S_f\rangle = \sum_{c,v,k} \frac{E_f^S}{E_c^{IPA} - E_v^{IPA}} A_{c,v,k}^S p_{cv,k}^{IPA} \qquad (10)$$

$$\langle S_m|p|S_f\rangle = \sum_{c_1,c_2,v_1,v_2,k} A_{c_1,v_1,k}^{S_m *} A_{c_2,v_2,k}^{S_f} \left( p_{c_1 c_2,k}^{IPA} \delta_{v_1 v_2} + p_{v_1 v_2,k}^{IPA} \delta_{c_1 c_2} \right) \qquad (11)$$

Finally, the $n$-photon absorption coefficient $\alpha_n$ is related to $W_n$ (eq.3) by

$$\alpha_n(\omega) = \frac{n\hbar\omega}{I^2} W_n(\omega) \quad (12)$$

, where $I$ is the applied light radiation intensity, which is related to $A_0$ by $A^2_0 = 2\pi cI/(n\omega^2)$, where $n$ is the refractive index and $c$ is the speed of light in vacuum. Hereafter, we use the notation $\alpha$ and $\beta$ for OPA and TPA coefficients, respectively.

## 3. Applications

### 3.1 Computational method

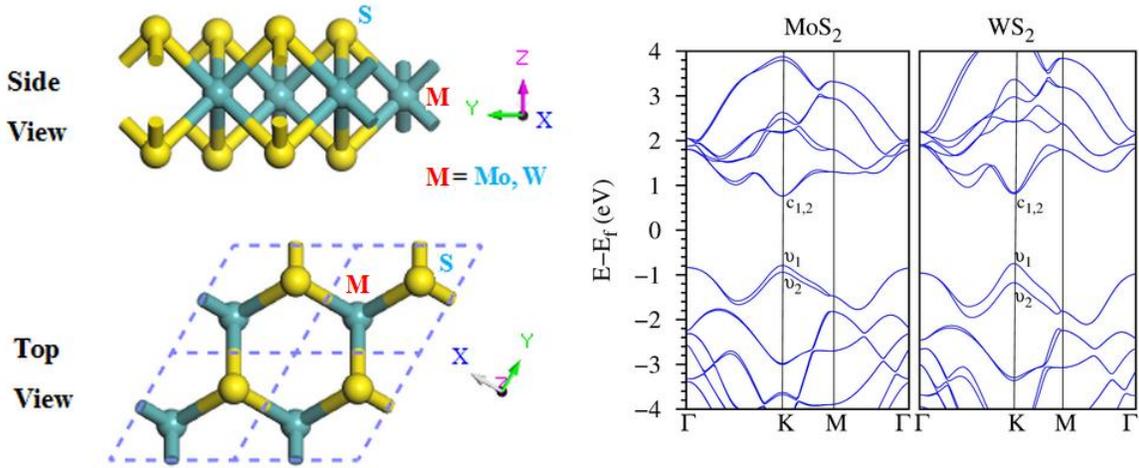

**Fig. 1. Crystal structures (left) and band structures (right) of MoS$_2$ and WS$_2$ monolayers. The dash lines (left) denote the unit cell.**

Crystal structures of MoS$_2$ and WS$_2$ monolayers are shown in Fig. 1. The unit cell with D$_{3h}$ symmetry was optimized by using the density functional theory within the GGA of PBE combined with the FP-LAPW method [13], as implemented in the ELK code [21]. A $k$-point mesh of 10×10×1, force threshold of 2×10$^{-4}$ a.u., and stress threshold of 1×10$^{-5}$ a.u. were used for optimizations. The relaxation of the unit cell was included in optimizations. A vacuum spacing larger than 15 Å was used to ensure negligible interaction between the slabs. The optimized structures were used for the band structure calculations. We performed the band structure calculations by using the GGA-PBE combined with the all-electron FP-LAPW method [13], as implemented in ELK code [21]. A $k$-point mesh of 18×18×1 was used for the band structure calculations. The spin-orbit coupling was considered. The optimized lattice parameters, electronic energy gaps, two transition energies at the K point, and the spin-orbit splitting energy are shown in table 1. The band structures and spin-orbit splitting energies are in good agreement with previous reports [5,22–24]. For instance, the $MoS_2$ monolayer is a direct gap semiconductor with an electronic energy gap ($E_g$) at the K point, where the conduction band minimum is doubly degenerate and the valence band maximum is split due to the spin-orbit coupling.

**Table 1. Optimized lattice parameters ($a = b$ in Å), electronic energy gaps ($E_g^{PBE}$ in eV), two transition energies ($E_{v1c}$ and $E_{v2c}$ in eV) at the K point, and the spin-orbit splitting energies ($\triangle_{so}$ in eV) based on the PBE calculation. $E_{sc}$ is the scissor correction value (eV) for the BSE calculation. The available experimental electronic energy gaps ($E_g^{Exp}$ in eV) are included.**

|        | $a = b$ | $E_g^{PBE}$ | $E_g^{Exp}$ | $E_{v1c}$ | $E_{v2c}$ | $\triangle_{so}$ | $E_{sc}$ |
|--------|---------|-------------|-------------------------------|-----------|-----------|------------------|----------|
| $MoS_2$ | 3.21   | 1.55        | 2.16 ± 0.04 [a], 2.15 or 2.35 [b] | 1.55      | 1.70      | 0.15             | 0.61     |
| $WS_2$  | 3.18   | 1.56        | 2.38 ± 0.06 [a], 2.47 [c]     | 1.56      | 1.99      | 0.43             | 0.82     |

[a] on fused quartz substrate with the STS measurement at room temperature [6,25]

[b] on highly ordered pyrolytic graphite substrate with the scanning tunneling spectroscopy (STS) measurement at 77 K [26]

[c] on monolayer graphene with the STS measurement at ~5 K [27]

For the optical properties, the energy band structures within the IPA were obtained by solving the self-consistent Kohn–Sham equations with the GGA-PBE functional, and the excitation states including electron-hole interaction were obtained by solving the BSE with a basis linearly expanded by the IPA states (Eq.7). The spin-orbit coupling was included in all calculations. The corresponding momentum matrix elements were also calculated by a homemade ELK interface which reads $p^{IPA}$ to calculate the BSE states based momentum matrix in terms of Eqs. 10 and 11. As the GGA-PBE calculation generally underestimates the band gap of solid, the scissor correction was used and the corresponding scissor value was given by the difference between the theoretical and experimental electronic energy gaps (table 1).

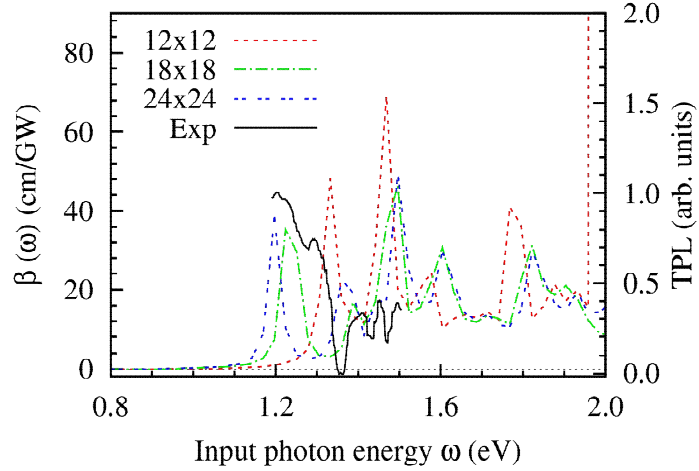

**Fig. 2. The *k*-dependence of the TPA spectra for WS$_2$ monolayer with a scissor value of 0.82 eV (table 1). The experimental TPA spectrum [28] is included for comparison.**

We calculated the TPA coefficients parallel to the monolayer plane ($\beta=\beta_{xx=yy}$, *x, y* directions defined in Fig. 1). Similar to the OPA calculation [4], we make a convergence test on the k-grid for the TPA calculation. Since the experimental TPA spectrum of WS$_2$ monolayer is available for comparison, we performed the test calculation on WS$_2$ monolayer. Figure 2 shows the *k*-dependence of *β* of WS$_2$ monolayer with a scissor value of 0.82 eV calculated by (2.38–1.56) (table 1). We observe a very similar distribution of absorption peaks for 18×18 and 24×24 k-grids and an agreement between the theoretical and experimental results (see a detailed analysis below). A smaller 12×12 k-grid leads to an absence of the first two-photon excitonic peak which appears in the experimental TPA spectrum [28]. Furthermore, as will be shown below, the OPA based on the 18×18 k-grid also agrees well with a very recent experimental OPA [6] in terms of peak positions and binding energy of excitonic states. It has been also suggested that the 18×18 k-grid yield the converged excitonic binding energy [4]. Thus, we will further discuss the TPA spectrum based on the excited states obtained from the 18×18 k-grid BSE calculations.

*3.2 Origin of TPA of MoS$_2$ and WS$_2$ monolayers*

Now, we further discuss the TPA spectrum of WS$_2$ monolayer. We also discuss the TPA spectrum of MoS$_2$ monolayer whose OPA including excitonic effect has been well studied [4,5,14,15]. Based on the convergence test above, we also use the 18×18 k-grid and the scissor value ($E_g^{Exp} - E_g^{PBE}$) to calculate the TPA spectrum of MoS$_2$ monolayer. Note that the difference in experimental conditions, such as substrate and temperature, leads to different electronic energy gaps. For consistence, we used the experimental electronic energy gaps based on the same experimental measurement [6,25] for MoS$_2$ and WS$_2$ monolayers (*i.e.*, 2.16 and 2.38 eV in table 1). Figure 3 shows the TPA spectra of MoS$_2$ and WS$_2$ monolayers based on the BSE and IPA calculations. We also show in Fig.3 the theoretical and experimental [6] OPA spectra to understand the TPA spectra. The experimental OPA spectra were obtained by Rigosi *et al.* [6] who used the optical reflectance contrast measurements at room temperature. They also reported the electronic band gap (table 1) based on the scanning tunneling spectroscopy measuement at room temperature. Note that we used this electronic band gap to make a scissor correction in the BSE calculations. As shown in Figs. 3c and 3d, our theoretical OPA are in good agreement with the experimental ones in terms of peak positions. Note that we have made a rigid shift of 0.05 and 0.10 eV for MoS$_2$ and WS$_2$ monolayers, respectively. These rigid shifts are valid because the experimental electronic band gaps used in our BSE calculations have uncertainty (table 1). And also note that the rigid shift mainly leads to the rigid shift of peak position and hardly affect the intensity of peak [29,30]. As shown in table 2, various theoretical and experimental methods yield very close transition energies for *A* and *B* excitons. However, very different electronic band gaps are reported in these theoretical and experimental works, which leads to different excitonic binding energies in the range of 0.2–1.0 eV [4–6,9,14,22,26,28,31–33]. Our present theoretical results are in good agreement with the theoretical [4] and experimental [6] ones.

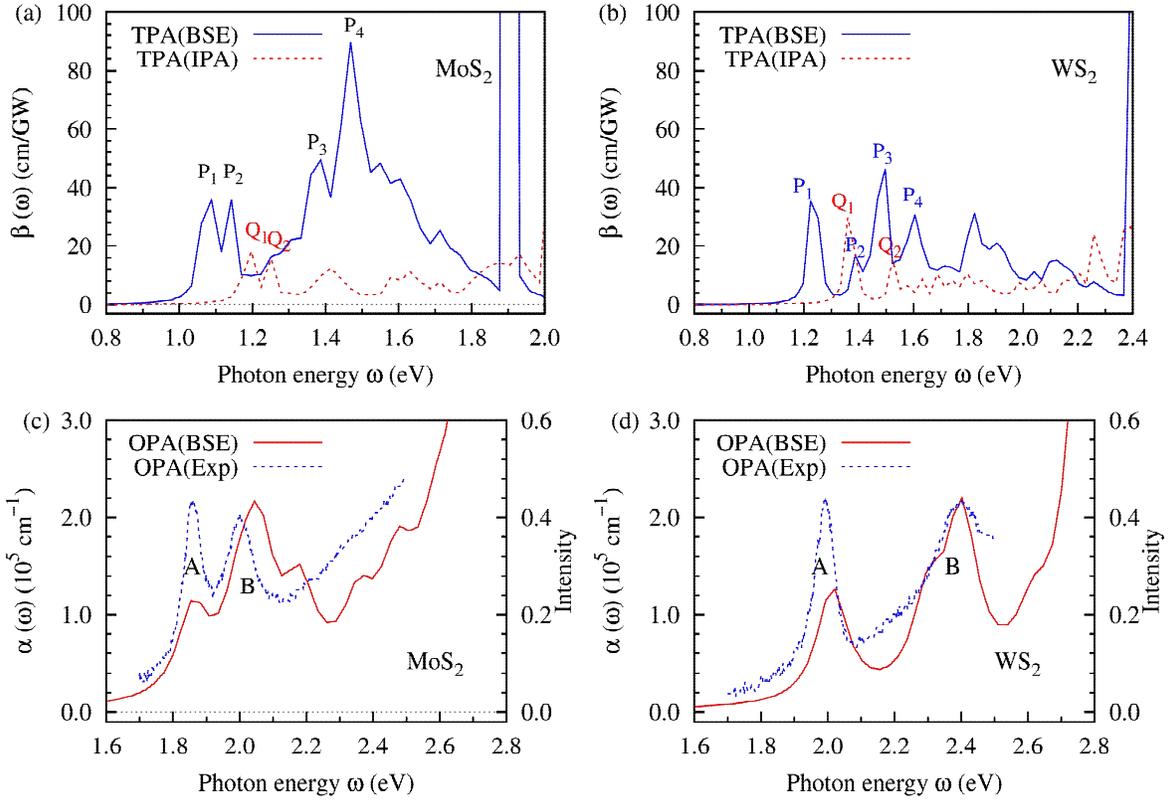

**Fig. 3.** TPA spectra of (a) $MoS_2$ and (b) $WS_2$ monolayers based on the BSE and IPA calculations. Theoretical and experimental [6] OPA spectra of (c) $MoS_2$ and (d) $WS_2$ monolayers.

Table 2. Positions (eV) of OPA and TPA peaks labeled in Fig. 3. $E_b$ is the binding energy of $A$ excitonic state (Figs.3c and 3d). Available experimental and theoretical results are included for comparison.

|  | OPA | | | TPA | | | | | |
|---|---|---|---|---|---|---|---|---|---|
|  | A | $E_b$ | B | $P_1$ | $P_2$ | $P_3$ | $P_4$ | $Q_1$ | $Q_2$ |
| $MoS_2$ | 1.91, 1.78[a], 1.88[b], 1.85[c], 1.89[d], 1.86[f] | 0.25, 0.31[f] | 2.09, 1.96[a], 2.03[b], 1.98[c], 2.03[d], 2.00[f] | 1.09 | 1.15 | 1.38 | 1.47 | 1.19 | 1.25 |
| $WS_2$ | 2.10, 1.84[a], 2.00[e], 2.02[f] | 0.28, 0.36[f] | 2.52, 2.28[a], 2.39[e], 2.40[f] | 1.22 | 1.36 | 1.46 | 1.58 | 1.36 | 1.53 |

[a] **GW-BSE calculation** [5]; [b] **absorption measurement** [34]; [c] **PL measurement** [35]; [d] **absorption measurement** [36]; [e] **absorption measurement** [37]; [f] **STS measurement** [6]

While the OPA spectra show the light exciton, the TPA spectra help us to identify the dark exciton [38] because TPA has different selection rule from OPA. To show the importance of excitonic effect in the TPA spectrum, we first consider the TPA spectrum based on the IPA calculation (IPA-TPA). As shown in Figs. 3a and 3b, the IPA-TPA spectra exhibit two distinct peaks (labeled by $Q_1$ and $Q_2$) near the onset of spectrum. To obtain an insight into the origin of these two peaks, we trace the SOS process and show in Fig. 4 the distribution of contributions from the k-points of the first Brillouin zone ($\sum_{vc}$ in Eq.5) used in the SOS calculation. We can see that the contributions mainly come from the k-points near six vertices of the first Brillouin zone such as K and K' points (see size of points). Furthermore, the energy difference between the $Q_1$ and $Q_2$ peaks is close to half the spin-orbit splitting energy. For instance, by tracing the SOS process, we find that $Q_1$ and $Q_2$ of $MoS_2$ monolayer can mainly attributed to the transition from $v_1$ to $c$ and from $v_2$ to $c$ (Fig. 1), respectively. At these k-points near the K point (Fig. 1), the spin-orbit splitting energy is 0.136 eV which is about twice the energy difference between $Q_1$ and $Q_2$ peaks (0.06 eV in table 2). Thus, the $Q_1$ and $Q_2$ peaks are associated with the spin-orbit coupling effect.

Then, we discuss the TPA spectra based on the BSE calculations (BSE-TPA). In Figs. 3a and 3b, we show the TPA spectra with the input photon energy below the OPA edge. Near the OPA edge, the TPA spectra possibly exhibit a strong OPA resonance due to the denominator of $(E^s - \hbar\omega)$ (Eq.9). For instance, the OPA edge of $MoS_2$ monolayer is located at 1.91 eV (table 2), which leads to a strong resonant enhancement of $\beta$ near 1.9 eV (Fig. 3a). Hereafter, we focus on the input photon energy below the OPA edge to avoid the OPA resonance. As shown in Fig. 3, there are distinct TPA excitonic peak with the input photon energy below the OPA edge. The selected characteristic peaks are labeled by $P_{1,2,3,4}$ and the corresponding photon energies are given in table 2. Likewise, to obtain an insight into the origin of these peaks, we identify the corresponding two-photon excitonic state by tracing the SOS process and show in Fig. 4 the distribution of weight ($\sum_{vc}|A^S_{vck}|^2$ in Eq.7) for the k-points of the first Brillouin zone used to construct the excitonic state in Eq.7. For the $P_1$ and $P_2$ peaks, similar to the $Q_1$ and $Q_2$ peaks, the contributions mainly come from the k-points near six vertices of the first Brillouin zone. However, compared to the $Q_1$ and $Q_2$ peaks, the $P_1$ and $P_2$ peaks have a ~0.1 eV red-shift (Fig. 3a and table 2), which suggests a significant excitonic effect. At the same time, the energy difference between the $P_1$ and $P_2$ peaks is 0.06 eV, which also imply these two peaks could be associated with the spin-orbit coupling effect. To demonstrate this view, we list in table 3 the $\sum_k|A^S_{vck}|^2$ (a summation on the k-points of the first Brillouin zone) for each $v \to c$ transition pair used in Eq.7 for $P_{1,2,3,4}$ peaks of $MoS_2$ monolayer. As shown in table 3, the transitions at the $P_1$ and $P_2$ peaks are dominated by the $v_1$

→ $c_2$ and $v_2$ → $c_1$ transition pairs, respectively. Note that in the vicinity of the K point, $c_1$ and $c_2$ are almost degenerate. Thus, as expected, the spin-orbit coupling leads to the $P_1$ and $P_2$ peaks. As for the $P_3$ and $P_4$ peaks, the contributions are mainly from the k-points around Γ point (Fig. 4), and the corresponding transitions are dominated by the transitions between ($v_1$, $v_2$) and ($c_1$, $c_2$) (table 3). Similar results can be obtained for $WS_2$ monolayer.

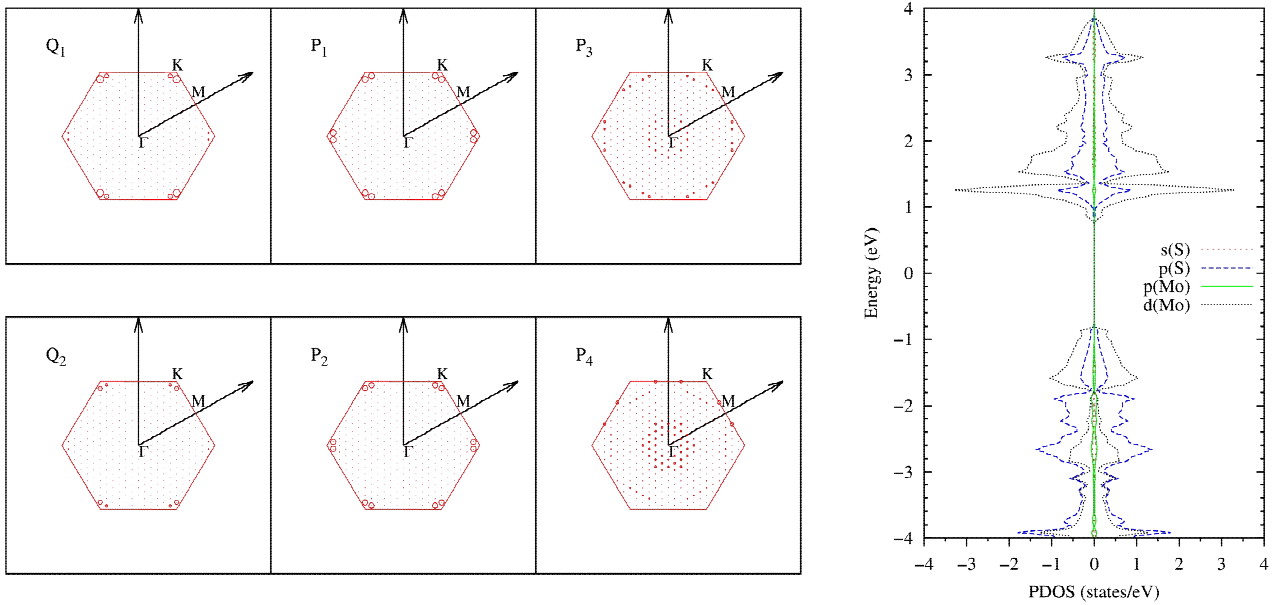

**Fig. 4.** (left) k-points of the first Brillouin zone used in $\sum_{vc}$ (Eq.5) and $\sum_{vc}|A^S_{vck}|^2$ (Eq.7) for $Q_{1,2}$ and $P_{1,2,3,4}$ peaks of $MoS_2$ monolayer, respectively. The size of point indicates the magnitude of contribution to summation for each k-point. The black arrows are the reciprocal vectors. (right) Projected density of states of $MoS_2$ monolayer.

**Table 3.** $\sum_k|A^S_{vck}|^2$ for each $v \to c$ transition pair in summation on the k-points of the first Brillouin zone for $P_{1,2,3,4}$ peaks of $MoS_2$ monolayer. For clarity, eight transition pairs (23, 24) → (27, 28, 29, 30) are not shown because their contributions to $\sum_k|A^S_{vck}|^2$ are very small (< 0.0001).

| v | c | $P_1$ | $P_2$ | $P_3$ | $P_4$ |
| --- | --- | --- | --- | --- | --- |
| 25 ($v_2$) | 27 ($c_1$) | 0.0 [a] | 0.9884 | 0.2136 | 0.1207 |
| 25 | 28 | 0.0 | 0.0 | 0.2482 | 0.2180 |
| 25 | 29 | 0.0 | 0.0 | 0.0109 | 0.0146 |
| 25 | 30 | 0.0 | 0.0 | 0.0338 | 0.0674 |
| 26 ($v_1$) [b] | 27 ($c_1$) [c] | 0.0 | 0.0 | 0.3445 | 0.3003 |
| 26 | 28 ($c_2$) | 0.9994 | 0.0094 | 0.1114 | 0.2083 |
| 26 | 29 | 0.0 | 0.0 | 0.0241 | 0.0514 |
| 26 | 30 | 0.0 | 0.0 | 0.0125 | 0.0179 |

[a] 0.0 means the value is small than 0.0001

[b] the highest valence band (Fig.1)

[c] the lowest conduction band (Fig.1)

Finally, to obtain an in-depth understanding of electronic transitions in the absorption spectrum, we show in Fig. 4 the projected density of states of $MoS_2$ monolayer. We can see that the valence band edge mainly consist of the $d$ orbital of Mo which are hybridized with the $p$ orbital of S, and that the $d$ orbital of Mo is mainly for the conduction band edge, in agreement with previous reports [9].

*3.3* Comparison with experiment

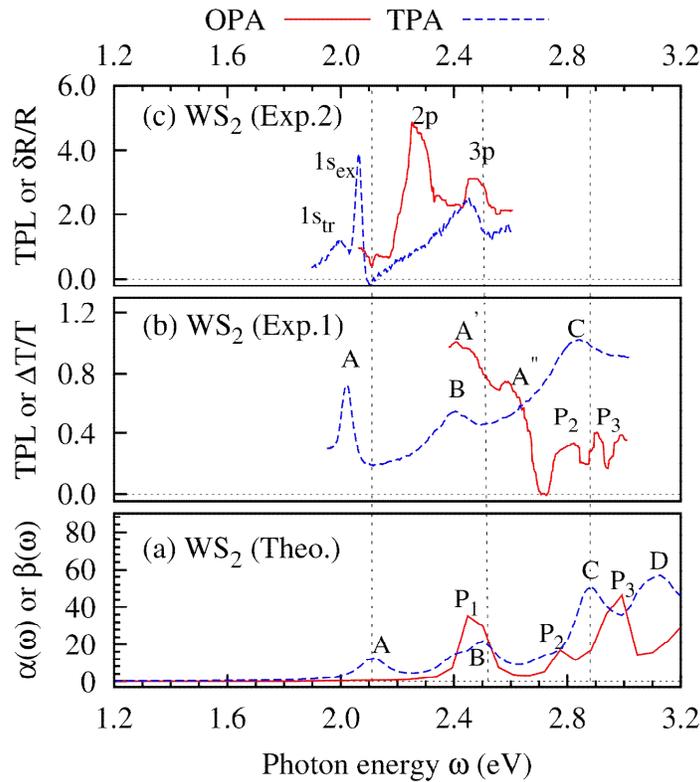

**Fig. 5. (a) theoretical (Theo.) and experimental (b) Exp.1 [28] and (c) Exp.2 [38] OPA and TPA spectra for WS$_2$ monolayer. For TPA, twice input photon energies (2ω) are used to plot. α(ω) and β(ω) are in $10^4$ cm$^{-1}$ and cm/GW units, respectively. TPL, δR/R, and ΔT/T indicate the two-photon luminescence, relative reflectance, and normalized differential transmission spectra, respectively. The vertical lines are for guiding the eyes.**

We make a comparison between theoretical and experimental OPA and TPA spectra for WS$_2$ monolayer. Figure 5 shows two experimental OPA and TPA spectra [28,38] and our theoretical ones for WS$_2$ monolayer. For OPA, as mentioned above, our theoretical A and B excitonic positions with a small rigid shift agree well with other theoretical or experimental results (Figs. 3c and 3d and table 2). Overall, Figure 5 also shows good agreements between different results. Note that in Exp.2, besides the A exciton, the negatively charged trion absorption peak at 2.0 eV was also detected [38], which is not found in Exp.1. In particular, there is an excellent agreement between the theoretical and experimental (Exp.1) spectra in terms of peak positions and line shape when a rigid shift of ~0.1 eV is made (also see Fig.3d above). However, as for TPA, there is no such a good agreement between different results. Due to different applied light ranges and other experimental conditions such as substrate and temperature, only a qualitative comparison to Exp.2 has been made by Zhu *et al.* [28] based on the two-dimensional hydrogen model. It was pointed out by Zhu *et al.* that the A' and A" peaks of TPA are likely assigned to the excited states of the A exciton. In our theoretical results, if we assign the P$_1$ peak to the excited state of the A exciton, the P$_2$ peak could be assigned to the excited state of the B exciton because the P$_1$ and P$_2$ peaks are associated with the spin-orbit splitting at the same k-points as the A and B excitonic peaks (see Fig.4 and Ref.[2]). For 2.7 eV < 2ω < 3.2 eV in Fig. 5a, the *alternative* peaks in OPA and TPA spectra indicate the importance of TPA in probing the excitonic dark states. Based on the position of C peak in OPA, we suggest that the P$_2$ and P$_3$ peaks in Exp.1 should be matched with those in our theoretical spectrum. Note that it is difficult to make a quantitative comparison between theoretical and experimental results because experimental measurements are usually performed on samples on substrate [28,38]. Even for experimental results, quantitative comparisons are also difficultly made owing to the difference in experimental conditions such as substrate and temperature. Thus, our theoretical OPA and TPA spectra are qualitatively well consistent with those in Exp.1.

## 4. Conclusions

We have presented a first-principles many-body formalism based on the BSE eigenstates for calculating the TPA spectrum of two-dimensional semiconductor materials. As applications, we have used this formalism to calculate the TPA spectra of $MoS_2$ and $WS_2$ monolayers. Compared to the IPA-TPA spectra, the BSE-TPA ones exhibit significant excitonic effects. By tracing the SOS process, we find that the first two BSE-TPA peaks on the onset of spectrum mainly originate from the transitions between the valance and conduction band edges at the vicinity of K point. At the higher applied photon energy, the two BSE-TPA peaks are dominated by the transitions at the k points around $\Gamma$ point. For $WS_2$ monolayer, the calculated BSE-TPA spectrum is in good agreement with the experimental one in terms of peak positions and line shape. Our theoretical BSE-TPA spectrum of $MoS_2$ is an important reference for experiments due to the similarity of electronic structures of $MoS_2$ and $WS_2$ monolayer.

## Acknowledgements

We appreciate the financial support from Natural Science Foundation of China Project 21303164 and the computational support from Zhejiang Key Laboratory for Reactive Chemistry on Solid Surfaces.

[36] Y. V. Morozov and M. Kuno, Appl. Phys. Lett. **107**, 083103 (2015).

[37] Y. Li, A. Chernikov, X. Zhang, A. Rigosi, H. M. Hill, A. M. van der Zande, D. A. Chenet, E.-M. Shih, J. Hone, and T. F. Heinz, Phys. Rev. B **90**, 205422 (2014).

[38] Z. Ye, T. Cao, K. O/'Brien, H. Zhu, X. Yin, Y. Wang, S. G. Louie, and X. Zhang, Nature **513**, 214 (2014).